\documentclass[12pt]{iopart}
\usepackage[dvips]{epsfig}

\def\##1{{\underline #1}}
\def\=#1{\underline{\underline #1}}
\def\*#1{{\bar#1}}

\def\eps{\epsilon}
\def\epso{\epsilon_0}
\def\muo{\mu_0}
\def\ko{k_0}
\def\lambdao{\lambda_0}
\def\etao{\eta_0}
\def\.{\mbox{ \tiny{$^\bullet$} }}

\def\aal{a_L}
\def\aar{a_R}
\def\bbl{r_L}
\def\bbr{r_R}
\def\ccl{t_L}
\def\ccr{t_R}

\def\le{\left(}
\def\ri{\right)}
\def\les{\left[}
\def\ris{\right]}

\def\up{\hat{\#{u}}_+}
\def\um{\hat{\#{u}}_-}
\def\ux{\hat{\#{u}}_x}
\def\uy{\hat{\#{u}}_y}
\def\uz{\hat{\#{u}}_z}

\def\r#1{(\ref{#1})}

\begin{document}

\title[Polarization--universal rejection by ambichiral structures]{Polarization--universal rejection
filtering by ambichiral structures made of indefinite dielectric--magnetic materials}

\author{Akhlesh Lakhtakia}
\address{Computational \& Theoretical Materials Sciences Group (CATMAS),
Department of Engineering Science \& Mechanics,
Pennsylvania State University, University Park, PA 16802--6812, USA}
\ead{akhlesh@psu.edu}




\begin{abstract}
An ambichiral structure comprising sheets of an anisotropic dielectric material rejects normally incident plane waves of
one circular polarization (CP) state but not of the other CP state,
in its fundamental Bragg regime. However, if the same structure is made of
an dielectric--magnetic material with indefinite  permittivity
and permeability dyadics,
it may function as a polarization--universal rejection filter because two of the four planewave components
of the electromagnetic field phasors in each sheet
are of the positive--phase--velocity type and two are of the negative--phase--velocity type.

\end{abstract}


\section{Introduction}
This communication combines two topics of recent interest in electromagnetics: (i) the ambichiral
structure that replicates the circular--polarization--sensitive filtering properties of cholesteric liquid crystals
and chiral sculptured thin films, and (ii) the dielectric--magnetic material with 
 indefinite  permittivity
dyadic and indefinite permeability dyadic.

The ambichiral structure is a structurally chiral pile of identical sheets that was conceived
by Reusch \cite{Reusch} to transmit normally incident
circularly polarized (CP)
plane waves of one handedness but reflect CP plane waves of the other handedness,
in a certain free--space--wavelength regime. This conceptualization influenced
early theoretical research on the optical response characteristics
of cholesteric liquid crystals \cite{JB,Jac92}. Following
a systematic study in 2004, Reusch's wavelength regime was identified
as merely the first of a potentially infinite number of Bragg regimes \cite{HLWDM}. For
optical applications, since then
the ambichiral structure has been experimentally realized  using 
sculptured--thin--film technology \cite{HLWDM,PBS2005}, and 
electro--optic versions have been suggested as electrically controlled
CP filters \cite{Lakh2006,DL2008}.

A   real symmetric dyadic is said to be indefinite if some of its eigenvalues are positive but the remaining ones are negative. Artificial materials with indefinite permittivity and
permeability dyadics came into prominence a few years ago, as such materials
can exhibit negative refraction \cite{PGLKT,SKS2004}. Due to hyperbolic,
instead of the usual elliptic, dispersion relations for
planewave propagation in these materials \cite{DIL2006A,LRSL}, several 
electromagnetic phenomenons~---~including surface--wave propagation
\cite{YSRK}, the 
Goos--H\"anchen shift \cite{XDW}, and diffraction by surface--relief gratings \cite{DLnjp,DIL2006B}~---~are exhibited by these materials in uncommon ways.

Motivated by these reports, an investigation was undertaken on the response
to a normally incident plane wave of an ambichiral structure made of
a dielectric--magnetic material with indefinite  permittivity
dyadic and indefinite permeability dyadic. True to expectation,
the usual CP--filtering response of ambichiral structures was not obtained.
Instead, a polarization--universal rejection response emerged,
indicating thereby the existence of a polarization--universal bandgap \cite{SZT,RL2007}.

The plan of this communication is as follows. Section~\ref{as} contains
a description of the ambichiral structure comprising 
orthorhombically anisotropic dielectric--magnetic
sheets. Section~\ref{reftrans} provides a succinct description of the boundary--value
problem to be solved in order to determine the response characteristics
of the ambichiral structure to a normally incident plane wave. Finally, numerical results are presented and discussed in Sec.~\ref{numres}.

A note on notation: Vectors are underlined and dyadics are double--underlined; the  cartesian unit vectors are represented
by $\ux$, $\uy$, and $\uz$; symbols
for column vectors and matrixes
are decorated by an overbar;  and an $\exp(-i\omega t)$ time--dependence is
implicit with $\omega$ as the angular frequency. The wavenumber, the wavelength, and the intrinsic impedance of free space are denoted by $\ko=\omega\sqrt{\epso\muo}$,
$\lambdao=2\pi/\ko$, and
$\etao=\sqrt{\muo/\epso}$, respectively, with $\muo$ and $\epso$ being  the permeability and permittivity of
free space.

\section{Ambichiral Structure}\label{as}

The ambichiral structure is a structurally chiral pile of $N$ identical  sheets, each of thickness
$D$ and infinite transverse extent. The $n^{th}$
sheet, $1\leq n\leq N$, occupies the region $(n-1)D< z< nD$;
thus, the total thickness of the pile is $L=ND$. The   sheets
are not necessarily electrically thin, and the structure can therefore be
considered as a 1--D photonic crystal \cite{DZ,Haus}.
The halfspaces $z \leq 0$ and
$z \geq L$ are vacuous.

The permittivity and permeability dyadics of the   $n^{th}$ sheet are chosen
as 
\begin{eqnarray}
\nonumber
&&\={\eps} (z) =\epso \,\={S}_{z}\left(h\xi_n\right)\cdot
\={S}_{y}\left(\chi\right)\cdot
\left(\epsilon_a\,\uz\uz+\epsilon_b\,\ux\ux+\epsilon_c\,\uy\uy\right)
\\[10pt]
&&\qquad\qquad\cdot\={S}_{y}\left(-\chi\right)\cdot\={S}_{z}\left(-\,h\xi_n\right)\,,\qquad (n-1)D< z< nD\,
\label{AAepsr}
\end{eqnarray}
and
\begin{eqnarray}
\nonumber
&&\={\mu} (z) =\muo \,\={S}_{z}\left(h\xi_n\right)\cdot
\={S}_{y}\left(\chi\right)\cdot
\left(\mu_a\,\uz\uz+\mu_b\,\ux\ux+\mu_c\,\uy\uy\right)
\\[10pt]
&&\qquad\qquad \cdot\={S}_{y}\left(-\chi\right)\cdot\={S}_{z}\left(-\,h\xi_n\right)\,,
\qquad (n-1)D< z< nD\,,
\label{AAmur}
\end{eqnarray}
respectively. Both dyadics indicate anisotropy of the orthorhombic
symmetry \cite{Lovett89,MLpo}.
The dyadic
\begin{equation}
\={S}_z(h\xi_n)= 
(\ux\ux+\uy\uy) \cos(h\xi_n) +(\uy\ux-\ux\uy)\sin(h\xi_n)+\uz\uz
\end{equation}
indicates rotation about the $z$ axis by an angle $h\xi_n$ with respect
to the first sheet, with 
\begin{equation}
\xi_n=(n-1)\frac{\pi}{q}\,,\qquad 1 \leq n \leq N\,,
\end{equation}
 the integer
$q\geq3$ \cite{HLWDM}, the ratio
$N/q$ an even integer, and
the parameter $h=1$ for structural right--handedness
and $h=-1$ for structural left--handedness. The number of
structural periods in the ambichiral structure is $N/2q$.
The dyadic 
\begin{equation}
\=S_y(\chi) = (\ux\ux+\uz\uz)\sin\chi + (\uz\ux-\ux\uz)\sin\chi +\uy\uy\,,\quad
\chi\in\left[0,\pi/2\right]\,,
\end{equation}
indicates a tilt with respect to the $xy$ plane by an angle $\chi$.

\section{Reflectances and Transmittances}\label{reftrans}

Suppose that an arbitrarily polarized plane wave is
normally incident on the ambichiral  structure
from the halfspace $z \leq 0$.
In consequence, a reflected plane wave must exist in  the same halfspace
and a transmitted plane wave in the halfspace $z \geq L$.
The electric field phasors associated with the two plane
waves in the halfspace $z\leq 0$ are stated as  
\begin{equation}
\#{E}_{inc}(\#{r}) =
 \le  \aal \, \up + \aar \, \um \ri \,
\exp( i \ko z ) \, , \quad z \leq 0\, ,
\end{equation}
and
\begin{equation}
\#{E}_{ref}(\#{r}) =
\le  \bbl \, \um + \bbr \, \up \ri \,
\exp( -i \ko z ) \, ,  \quad z \leq 0\, ,
\end{equation}
where $\#u_\pm = (\ux \pm i \uy)/\sqrt{2}$. Likewise,
the electric field phasor in the halfspace $z\geq L$ is
represented as
\begin{equation}
\#{E}_{trs}(\#{r}) =
 \le  \ccl \, \up + \ccr \, \um \ri \,
\exp\les i \ko (z -L)\ris  \,, \quad  z \geq L \, .
\end{equation} 
Here, $\aal$ and $\aar$ are the known amplitudes
of the left-- and the right--CP (LCP \& RCP)
components
of the incident plane wave;
$\bbl$ and $\bbr$ are the unknown amplitudes
of the reflected plane wave components; while $\ccl$ and $\ccr$
are the unknown amplitudes
of the transmitted plane wave components.  

The procedure to obtain the unknown reflection and transmission amplitudes
involves the following 4$\times$4
matrix relation \cite[Chap. 10]{LMbook}: 
\begin{equation}
\label{eq8}
\*f_{exit} = \*M\cdot\*f_{entry}\,.
\end{equation}
The column vectors
\begin{equation}
\label{eq9}
\*f_{entry} = \frac{1}{\sqrt{2}}\, \left(
\begin{array}{cc}
(\bbl+\bbr) + (\aal+\aar)\\ i\les -(\bbl-\bbr) + (\aal-\aar)\ris\\
-i\les (\bbl-\bbr) + (\aal-\aar)\ris/\etao\\
-\les (\bbl+\bbr) - (\aal+\aar)\ris/\etao \end{array}\right) \,
\end{equation}
and
\begin{equation}
\*f_{exit} = \frac{1}{\sqrt{2}}\, \left(
\begin{array}{cc}
 \ccl+\ccr\\ i\ (\ccl-\ccr) \\
-i (\ccl-\ccr)/\etao\\
(\ccl+\ccr) /\etao \end{array}\right)\,
\end{equation}
derive from the tangential components of the electric and magnetic field phasors
 at the entry and the exit
pupils, respectively.
The 4$\times$4 matrix \cite{Lakh2006}
\begin{equation}
\label{meqn}
\*M =  \exp \Big( i\*P_ND\Big)\cdot\exp \Big( i\*P_{N-1}D\Big)\cdot\dots\cdot
\exp \Big( i\*P_2D\Big)\cdot\exp \Big( i\*P_1D\Big)\,,
\end{equation}
encapsulating the planewave response of the entire ambichiral structure,
contains the matrix
\begin{equation}
\*P_n= \*B_n\cdot\*P_1\cdot \*B_n^{-1}\,,\qquad 1\leq n\leq N\,,
\end{equation}
where
\begin{equation}
\*B_n=\left(\begin{array}{cccc}
\cos(h\xi_n)&-\,\sin(h\xi_n)&0&0\\
\sin(h\xi_n) &\cos(h\xi_n) & 0 & 0\\
0&0&\cos(h\xi_n)&-\,\sin(h\xi_n)\\
0&0&\sin(h\xi_n)&\cos(h\xi_n)
\end{array}\right)\, , \quad 1\leq n\leq N\,,
\end{equation}
and the matrix \cite{WL1993}
\begin{equation}
\*P_1=\left(
\begin{array}{cccc}
0 & 0 & 0 &\omega\muo\mu_c\\
0& 0 &-\omega\muo\mu_d & 0\\
0 &-\omega\epso\epsilon_c&0&0\\
\omega\epso\epsilon_d&0&0&0
\end{array}
\right)\,
\end{equation}
involves
\begin{eqnarray}
&&\epsilon_d=\epsilon_a\epsilon_b/\left(\epsilon_a\cos^2\chi+\epsilon_b\sin^2\chi\right)\,,
\\
&&\mu_d=\mu_a\mu_b/\left(\mu_a\cos^2\chi+\mu_b\sin^2\chi\right)\,.
\end{eqnarray}

The reflection amplitudes $r_{L,R}$ and the transmission
amplitudes $t_{L,R}$  can be computed for specified incidence amplitudes
($\aal$ and $\aar$) by solving \r{eq8}. Interest usually lies  in determining
the reflection and transmission coefficients
entering the 2$\times$2 matrixes in the following two relations:
\begin{eqnarray}
\label{eq15}
\left( \begin{array}{cccc} \bbl \\ \bbr  \end{array}\right)  &=&
\left( \begin{array}{cccc} r_{LL} & r_{LR} \\ r_{RL} & r_{RR}\end{array}\right) \,
\left( \begin{array}{cccc} \aal \\ \aar  \end{array}\right) \, , \\
\label{eq16}
\left( \begin{array}{cccc} \ccl \\ \ccr  \end{array}\right)  &=&
\left( \begin{array}{cccc} t_{LL} & t_{LR} \\ t_{RL} & t_{RR}\end{array}\right) \,
\left( \begin{array}{cccc} \aal \\ \aar  \end{array}\right)
\, .
\end{eqnarray}
Both  2$\times$2 matrixes are defined phenomenologically.
The co--polarized transmission coefficients are denoted by $t_{LL}$ and
$t_{RR}$,
and the cross--polarized ones by $t_{LR}$ and $t_{RL}$; and similarly for the
reflection coefficients in \r{eq15}.
Reflectances and transmittances are denoted, e.g., as
$T_{LR} = |t_{LR}|^2$.

\section{Numerical Results and Discussion}\label{numres}
Let us assume that chosen material has negligible dispersion and
dissipation over the free--space--wavelength range of interest, for the sake
of simplicity.
With the assumption that both the relative permittivity and the relative
permeability dyadics are positive definite (i.e., $\epsilon_{a,b,c}>0$
and $\mu_{a,b,c}>0$),  the fundamental Bragg regime of the ambichiral
structure for normally incident plane waves has \cite{HLWDM}
\begin{equation}
\label{eqBr}
\lambdao^{Br}= qD \left(\sqrt{\epsilon_c\mu_d}+\sqrt{\epsilon_d\mu_c}\right)
\end{equation}
as its center--wavelength. This equation must also hold
if both the relative permittivity and the relative
permeability dyadics are negative definite (i.e., $\epsilon_{a,b,c}<0$
and $\mu_{a,b,c}<0$) \cite{Lakh-am,Lakh-oe}.
Even when the two dyadics are indefinite, we must ensure that the
ambichiral structure is electromagnetically penetrable by normally incident plane waves, in general. Therefore let us impose
the twin restrictions \cite{DLnjp,DIL2006B}
\begin{equation}
\label{restrictions}
\left.\begin{array}{l}
\epsilon_c\,\mu_d >0\\
\epsilon_d\,\mu_c>0
\end{array}\right\}\,.
\end{equation}

The following four cases were chosen for numerical investigation:
\begin{itemize}
\item PosDef: $\epsilon_c=2.7$, $\epsilon_d=3.2$, $\mu_c=1.1$,
$\mu_d=1.2$.

\item NegDef: $\epsilon_c=-2.7$, $\epsilon_d=-3.2$, $\mu_c=-1.1$,
$\mu_d=-1.2$.

\item Indef-1: $\epsilon_c=-2.7$, $\epsilon_d=3.2$, $\mu_c=1.1$,
$\mu_d=-1.2$.

\item Indef-2: $\epsilon_c=2.7$, $\epsilon_d=-3.2$, $\mu_c=-1.1$,
$\mu_d=1.2$.

\end{itemize}
All four cases satisfy the restrictions \r{restrictions}.

\begin{figure}[!ht]
\centering \psfull
\epsfig{file=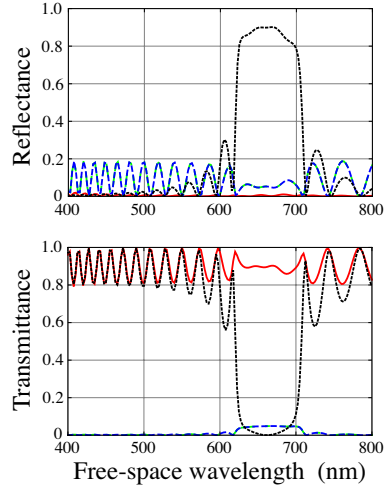, width=5cm}
\caption{\sf Case PosDef:
Reflectances and transmittances of an ambichiral structure
as functions of the free--space wavelength $\lambdao$. 
The following structural parameters were used for these
plots: $h=1$, $q=3$, $N=20q$, and $qD=200$~nm. The
constitutive parameters are as follows: $\epsilon_c=2.7$, $\epsilon_d=3.2$, $\mu_c=1.1$, and
$\mu_d=1.2$. Solid red lines are for $R_{LL}$ and $T_{LL}$,
black dotted lines for $R_{RR}$ and $T_{RR}$, blue
dashed lines for $R_{RL}$ and $T_{RL}$, and green
dash--dotted lines for $R_{LR}$ and $T_{LR}$.
Interchange the subscripts $L$ and $R$ in the reflectances
and transmittances for either (i) $h=-1$ \emph{or} (ii) Case NegDef
($\epsilon_c=-2.7$, $\epsilon_d=-3.2$, $\mu_c=-1.1$,
$\mu_d=-1.2$). The subscripts $L$ and $R$ must not be
interchanged if both (i) \emph{and} (ii) hold together.
\label{FigPosDef}
}
\end{figure}

Computed spectrums of the eight reflectances and transmittances
are shown in Fig.~\ref{FigPosDef} for Case PosDef and the
following structural parameters: $h=1$, $q=3$, $N=20q$, and $qD=200$~nm. 
The fundamental Bragg regime in the
figure as a high--reflectance feature for incident RCP
plane waves is correctly predicted by \r{eqBr}. In the same free--space--wavelength
regime, the transmission of incident 
LCP plane waves is very high. This CP--discriminatory phenomenon
is called the circular Bragg phenomenon \cite{LMbook}. It occurs when
the number of structural periods   is sufficiently large, the sufficiency
depending on the magnitude of the difference
$ \left(\sqrt{\epsilon_c\mu_d}-\sqrt{\epsilon_d\mu_c}\right)$, which
quantity may be called an effective local linear birefringence. The larger
this birefringence in magnitude, the fewer the structural periods required to
observe (and exploit) the circular Bragg phenomenon.

 The reflectances and transmittances for RCP and
LCP plane waves in Fig.~\ref{FigPosDef}
 shall have to be interchanged if either \begin{itemize}
\item[(i)] the sign of $h$ is changed,
or
\item[(ii)] the constitutive parameters for Case NegDef were to be used
instead of those for Case PosDef.
\end{itemize}
The \emph {equivalence} of the changes (i) and (ii) has been established
analytically \cite{Lakh-oe}:
When the real parts of the permittivity and the permeability
dyadics of a structurally chiral, magnetic--dielectric material are
reversed in sign, the circular Bragg phenomenon displayed by the material
in terms of reflectances and transmittances
suffers a change which indicates that the structural
handedness has been, in effect, reversed; additionally, the reflection
and transmission coefficients suffer phase reversal. If both changes (i)
and (ii) are required to occur simultaneously, they cancel out each other's
individual effects.

\begin{figure}[!ht]
\centering \psfull
\epsfig{file=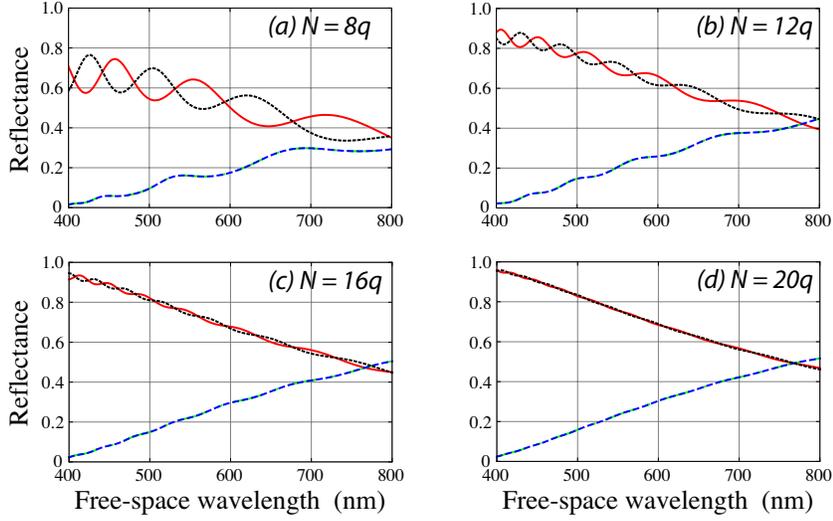, width=11cm}
\caption{\sf Case Indef-1:
Reflectances of an ambichiral structure
as functions of the free--space wavelength $\lambdao$. 
The following structural parameters were used for these
plots: $h=1$, $q=3$, and $qD=200$~nm. The
constitutive parameters are as follows: $\epsilon_c=-2.7$, $\epsilon_d=3.2$, $\mu_c=1.1$, and
$\mu_d=-1.2$. Solid red lines are for $R_{LL}$,
black dotted lines for $R_{RR}$, blue
dashed lines for $R_{RL}$, and green
dash--dotted lines for $R_{LR}$.
(a) $N=8q$, (b) $N=12q$, (c) $N=16q$, (d) $N=20q$.
\label{FigIndef1R}
}
\end{figure}

\begin{figure}[!ht]
\centering \psfull
\epsfig{file=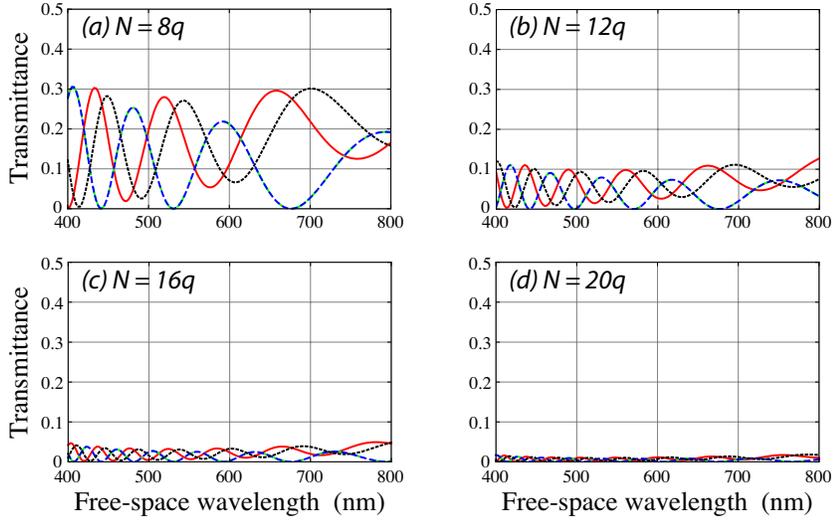, width=11cm}
\caption{\sf Case Indef-1: Same as Fig.~\ref{FigIndef1R}, except that
the spectrums of the four transmittances are plotted.
Solid red lines are for $T_{LL}$,
black dotted lines for $T_{RR}$, blue
dashed lines for $T_{RL}$, and green
dash--dotted lines for $T_{LR}$. 
\label{FigIndef1T}
}
\end{figure}

The situation changes completely for Case Indef-1. Spectrums of the
four reflectances are plotted in Fig.~\ref{FigIndef1R}, and those
of the four transmittances in Fig.~\ref{FigIndef1T}, for
$h=1$, $q=3$,  and $qD=200$~nm. These spectrums are provided
for ambichiral structures with 4, 6, 8, and 10 structural periods.
When the number of structural periods is small, the responses
to incident RCP and LCP plane waves are different. As the number
of structural periods increases, the discrimination between the responses
to CP plane waves of different handednesses decreases
and virtually vanishes for $N=20q$ in the two figures; concurrently,
the transmittances become increasingly smaller.

Qualitatively
similar conclusions were drawn from the spectrums of the reflectances
and transmittances for Case Indef-2, for which reason those
spectrums have not been provided here. Furthermore, for both Cases Indef-1
and Indef-2, the same conclusions emerged for all values of $q\leq 50$.

\def\etadc{\eta_{dc}}
\def\etacd{\eta_{cd}}
\def\kdc{k_{dc}}
\def\kcd{k_{cd}}

Taken together, the foregoing results offer the following
significant result: an ambichiral structure,
made of
an dielectric--magnetic material with indefinite  permittivity
dyadic and indefinite permeability dyadic and containing a sufficiently
large number of structural periods,
can function as a polarization--universal rejection filter in its fundamental Bragg
regime.  
Despite its structural chirality, an ambichiral structure thus does
not necessarily function as a CP--discriminatory filter.  

The polarization--universal rejection in Figs.~\ref{FigIndef1R}
and \ref{FigIndef1T} occurs over a much larger bandwidth than the
CP--discriminatory rejection in Fig.~\ref{FigPosDef}. However,
this observation is subject to modification when both dissipation
and dispersion are considered.

In order to understand the different response characteristics for
Cases PosDef and NegDef on the one hand and Cases Indef-1 and
Indef-2 on the other, the electromagnetic field phasors inside
the ambichiral structure have to be examined. Since all sheets are
identical, except for a rotation about the $z$ axis, it suffices to examine the
fields in the sheet labeled $n=1$. 
In this sheet, the electromagnetic
field phasors may be represented in terms of four plane waves as
\begin{equation}
\left.\begin{array}{l}
\#E(\#r)=- \etao\etacd\left(A^{(-)}\,e^{-i\ko\kcd z}-A^{(+)}\,e^{i\ko\kcd z}\right)\ux\\
\qquad +\,\etao\etadc\left(B^{(-)}\,e^{-i\ko\kdc z}-B^{(+)}\,e^{i\ko\kdc z}\right)\uy
\end{array}\right\}\,,\quad 0\leq z \leq D\,,
\end{equation}
and
\begin{equation}
\left.\begin{array}{l}
\#H(\#r)=\left(A^{(-)}\,e^{-i\ko\kcd z}+A^{(+)}\,e^{i\ko\kcd z}\right)\uy\\
\qquad + \,\left(B^{(-)}\,e^{-i\ko\kdc z}+B^{(+)}\,e^{i\ko\kdc z}\right)\ux
\end{array}\right\}\,,\quad 0\leq z \leq D\,,
\end{equation}
where $A^{(\pm)}$ and $B^{(\pm)}$ are coefficients of expansion, and
\begin{equation}
\left.\begin{array}{ll}
\etacd=\frac{\sqrt{\mu_c}}{\sqrt{\epsilon_d}}\,, &
\kcd= {\sqrt{\mu_c}}\,{\sqrt{\epsilon_d}}\\[5pt]
\etadc=\frac{\sqrt{\mu_d}}{\sqrt{\epsilon_c}}\,, &
\kdc= {\sqrt{\mu_d}}\,{\sqrt{\epsilon_c}}
\end{array}\right\}\,.
\end{equation}
The projection of the wave vector $\mp \ko\kcd\,\uz$ on the time--averaged Poynting vector 
$\mp\etao\etacd\,\vert A^{(\mp)}\vert^2\,\uz$
of the planewave component associated with $A^{(\mp)}$  
is either  (i) positive
if $\mu_c>0$ or (ii) negative if $\mu_c<0$.  Likewise,
the projection of the wave vector $\mp \ko\kdc\,\uz$ on the time--averaged Poynting vector 
$\mp\etao\etadc\,\vert B^{(\mp)}\vert^2\,\uz$
of the planewave component associated with $B^{(\mp)}$  
is either (i) positive
if $\mu_d>0$ or (ii) negative if $\mu_d<0$.  

For Case PosDef, all four planewave components inside each sheet are
of the positive--phase--velocity (PPV) type. For Case NegDef,
all four are of the negative--phase--velocity (NPV) type \cite{MLpre}.
For either Case Indef-1 or Indef-2, two planewave components
are of the NPV type and two of the PPV type. Both theory 
and experiment \cite{Reusch}--\cite{DL2008} show that CP--discriminatory
rejection is possible
in the fundamental Bragg regime when all four
planewave components are of the PPV type. From the conjugate
invariance of the frequency--domain Maxwell equations
\cite{Lakh-motl}, it follows
that CP--filtering must be possible
in the fundamental Bragg regime when all four
planewave components are of the NPV type. The polarization--universal
rejection exemplified by Figs.~\ref{FigIndef1R} and \ref{FigIndef1T}
for Cases Indef-1 and Indef-2~---~despite the pile
of sheets being structurally chiral~---~must therefore
be attributed to the fact that two planewave components out of four are
of the PPV type and  two of the NPV type.

\vskip 1cm
\noindent{\bf Acknowledgments.} This communication is dedicated to the fearless
campaign of Iftikhar Choudhury, Aitzaz Ahsan and others in defense
of democracy, justice, and constitutional rule of law.

\end{document}